\documentclass[10pt]{article}

\usepackage{graphicx}
\usepackage{caption} 
\usepackage{cite}
\usepackage[utf8]{inputenc}
\usepackage[T1]{fontenc}
\usepackage{authblk}
\usepackage{hyperref} 
\usepackage{xcolor} 
\usepackage{amsmath}
\usepackage{booktabs}

\usepackage[left=25mm,top=25mm,bottom=25mm,right=25mm]{geometry}
\title{
{\bf On Modeling Anisotropic Quark Stars: The Role of Anisotropy in Radial Oscillation Spectra}
}

\author{Grigoris Panotopoulos\thanks{grigorios.panotopoulos@ufrontera.cl}}

\affil{Departamento de Ciencias F{\'i}sicas, Universidad de La Frontera, Casilla 54-D, 4811186 Temuco, Chile.}

\begin{document}

\date{}

\maketitle

\begin{abstract}
We model the compact object Cen X-3, which is considered to be a good strange quark star candidate of known mass and radius, incorporating a negative anisotropic factor, and we compute the frequencies of the ten lowest radial oscillation modes. We introduce the anisotropy in three different manners and investigate its impact on the spectra.
\end{abstract}

\bigskip

\noindent  {\bf Keywords:} Relativistic stars; Anisotropic fluid spheres; Asteroseismology; Radial oscillations.

\bigskip

%%%%%%%%%%%%%%%%%%%%%%%
\section{Introduction}
%%%%%%%%%%%%%%%%%%%%%%%

Compact stars \cite{Shapiro:1983du, Sedrakian:2006mq} provide a unique theoretical and observational laboratory for studying the behavior of matter under extreme conditions of density, pressure and gravitational fields. In particular, neutron stars \cite{Lattimer, Ozel} and quark stars \cite{Weber} probe regions of the Quantum Chromodynamics (QCD) phase diagram inaccessible to terrestrial experiments, while at the same time offering an arena to test relativistic gravity in the strong-field regime. Recent advances in multi-messenger Astronomy, including high-precision pulsar timing, X-ray observations and gravitational-wave detections from binary mergers, have considerably improved the observational constraints on the masses, radii and tidal properties of compact objects. These developments have renewed the interest in realistic modeling of relativistic stellar configurations and their dynamical stability properties.

\smallskip

The standard description of compact stars usually assumes that matter behaves as an isotropic perfect fluid, for which the radial and tangential pressures coincide. However, several physical mechanisms involving relativistic particles may naturally induce anisotropic stresses at supranuclear densities \cite{Ruderman:1972aj}, such as phase transitions \cite{aniso3}, pion condensation \cite{aniso4}, or in presence of type 3A super-fluid \cite{aniso5}. In the presence of anisotropy, the radial pressure, $p_r$, differs from the tangential pressure, $p_t$, and the anisotropic factor,
$\Pi(r)=p_t(r)-p_r(r)$, modifies the hydrostatic equilibrium equation through an additional force term. Depending on its sign and magnitude, anisotropy may significantly alter the mass-radius relation, compactness, red-shift and stability properties of compact stars.

\smallskip

An important aspect in the study of relativistic stars is their response to small perturbations. Radial oscillations of pulsating stars constitute one of the most direct probes of dynamical stability, since the sign of the squared fundamental frequency determines whether the equilibrium configuration is stable or unstable against infinitesimal radial perturbations. Furthermore, the oscillation spectrum carries valuable information regarding the internal composition and equation of state of the stellar matter \cite{Li}. The study of radial modes therefore provides a complementary diagnostic tool to equilibrium observables, such as masses and radii. In recent years, the possibility of extracting oscillation signatures from gravitational-wave observations has further motivated detailed analyses of stellar pulsations in increasingly realistic compact-star models.

\smallskip

In anisotropic stellar systems, the oscillation spectrum is affected not only by the equation-of-state but also by the specific prescription adopted for the anisotropic sector. Since the microscopic origin of anisotropy in ultradense matter remains uncertain, most investigations rely on phenomenological models for the anisotropic factor. Consequently, it becomes relevant to examine how different anisotropy prescriptions influence the radial stability and pulsation spectra of compact stars with the same global astrophysical properties.

\smallskip

In the present work we investigate the radial oscillation spectra of anisotropic compact stars by considering a stellar configuration compatible with observational constraints on the mass and radius of a known compact object. We model the anisotropy using three distinct approaches. First, we consider the framework based on Herrera's vanishing complexity condition \cite{herrera}, where the anisotropy is determined through the requirement that the complexity factor of the system vanishes. This prescription establishes a non-trivial relation between the anisotropy and the density gradient, leading to physically motivated interior configurations. Second, we adopt a phenomenological anisotropy model proportional to the local compactness of the configuration, which effectively couples the anisotropic stresses to the strength of the gravitational field inside the star. Third, we construct an analytical stellar solution by prescribing a suitable profile for the mass function, from which the thermodynamic variables and anisotropy are obtained consistently through the Einstein field equations.

\smallskip

For each anisotropy model we solve the equilibrium structure equations and subsequently analyze the corresponding radial pulsation modes within the framework of linear perturbation theory. Particular attention is devoted to the dependence of the fundamental and higher-order oscillation frequencies on the anisotropy prescription, as well as to the resulting stability criteria. By comparing models reproducing the same observed mass and radius, we aim to isolate the role played by anisotropy in determining the dynamical response of compact stars. The differences between isotropic and anisotropic stars has been studied in previous publications, see e.g. \cite{Greg1,Greg2}, and therefore in this study we shall only consider stars made of anisotropic matter, with the purpose to investigate the role of anisotropy in radial oscillation spectra. Moreover, previous studies on the topic published by the author and co-workers, such as Refs. [12,42], were focused on structural properties of relativistic stars; the oscillation spectra were not computed. In addition to that, the assumed equations-of-state were different.

\smallskip

This work is organized as follows. In Sec. II we present the general formalism for anisotropic relativistic stars and summarize the equilibrium equations governing the stellar structure. In Sec. III we introduce the three anisotropy models considered in this study. In Sec. IV we develop the equations describing radial perturbations and the associated Sturm-Liouville eigenvalue problem. In Sec. V we present and discuss the numerical results for the equilibrium configurations and oscillation spectra. Finally, Sec. VI contains our conclusions and perspectives for future work. We adopt the mostly positive metric signature, and we work in geometric units where $G=1=c$.

%%%%%%%%%%%%%%%%%%%%%%%%%%%%%%%%%%%%%%%%%%%%%%%%%%
\section{Structure equations of anisotropic stars}
%%%%%%%%%%%%%%%%%%%%%%%%%%%%%%%%%%%%%%%%%%%%%%%%%%

We work with relativistic fluid spheres in four dimensions, with a vanishing cosmological constant, and assuming non-rotating objects. Here we briefly review the structure equations for interior stellar solutions, starting from the field equations of Einstein's General Relativity \cite{Einstein:1915ca}
\begin{equation}
    G_{mn} \equiv R_{mn} - \frac{1}{2} \: R \: g_{mn} = 8 \pi  T_{mn},
\end{equation}

\medskip

\noindent where $T_{mn}$ is the energy-momentum tensor of the matter content, $g_{mn}$ is the metric tensor, $R_{mn}$ and $R$ are the Ricci tensor and Ricci scalar, respectively, while $G_{mn}$ is the Einstein tensor.

\smallskip

The line element in Schwarzschild-like coordinates $\{ t, r, \theta, \phi \}$ for static, spherically symmetric geometries is given by
\begin{equation}
    d s^2 = - e^{\nu} d t^2 + e^{\lambda} d r^2 + r^2 (d \theta^2 + \sin^2 \theta \: d \phi^2) , 
\end{equation}

\medskip

\noindent while the anisotropic matter content viewed as a perfect fluid is described by a stress-energy tensor of the form
\begin{equation}
T_a^b = Diag(-\rho, p_r, p_t, p_t),
\end{equation}

\medskip

\noindent where $p_r$ is the radial pressure of the fluid, $\rho$ is its energy density, and $p_t$ is the tangential pressure. Depending on the matter content, $p_r$ and $\rho$ satisfy a certain EoS $p(\rho)$.

\smallskip

To compute the stellar mass $M$ and radius $R$ we need to integrate the Tolman-Oppenheimer-Volkoff (TOV) equations \cite{Oppenheimer:1939ne, Tolman:1939jz} to obtain interior solutions of the star describing hydrostatic equilibrium. The TOV equations, for just one fluid component, are given by
\begin{align}
    \displaystyle m'(r) &= 4 \pi r^2 \rho (r) , \\ 
    \displaystyle \nu' (r) &= 2 \: \frac{m(r) + 4 \pi r^3 p_r(r)}{ r^2 \left( 1 - 2 m(r) / r \right) } , \\
    \displaystyle p_r'(r) &= - [ \rho(r) + p_r(r) ] \; \frac{\nu' (r)}{2} + \frac{2 \: \Pi (r)}{r} ,
\end{align}

\smallskip

\noindent where the prime denotes differentiation with respect to the radial coordinate $r$, the mass function $m(r)$ is defined by
\begin{equation}
    \displaystyle e^{\lambda} = \frac{1}{1 - \frac{2 \: m(r)}{r}} .
    \label{eq:2}
\end{equation}
and $\Pi(r)=p_t-p_r$ is the anisotropic factor. The third TOV equation is equivalent to the stress-energy conservation, $\nabla_\mu T^{\mu \nu}=0$, which is a consequence of the Bianchi identity, $\nabla_\mu G^{\mu \nu}=0$, with $\nabla_\mu$ being the covariant derivative.

\smallskip

To solve the system of coupled differential equations, we integrate them throughout the star imposing initial conditions at the center ($r = 0$), and matching conditions at the surface of the star ($r = R$). 
In particular, the initial conditions at the origin are as follows 
\begin{equation}
    m(0) = 0 ,
\end{equation}
\begin{equation}
    p_r(0) = p_c ,
\end{equation}

\medskip

\noindent with $p_{c}$ being the central value of the radial pressure, while upon comparison to the exterior vacuum solution given by the Schwarzschild geometry \cite{Schwarzschild:1916uq}
\begin{equation}
d s^2 = -f(r) d t^2 + f(r)^{-1} dr^2 + r^2 (d \theta^2 + \sin^2 \theta \: d \phi^2), \; \; \; \; \; \; \; f(r) = 1-2 \frac{M}{r} .
\end{equation}

Since the radial pressure is zero outside the star, and due to the continuity of the two metric potentials at $r=R$, the matching conditions at the surface of the star yield
\begin{equation}
p_r(R) = 0,
\end{equation}
\begin{equation}
m(R) = M, 
\end{equation}
\begin{equation}
e^{\nu(R)} = 1-2 \frac{M}{R} .
\end{equation}

\medskip

The first two conditions allow us to compute the radius and the mass of the star, and finally the other metric potential, $\nu(r)$, may be computed by
\begin{equation}
    \displaystyle \nu (r) = \ln \left( 1 - \frac{2 M}{R} \right) + 2 \int_R^r \frac{m(x) + 4 \pi x^3 p_r(x)}{ x^2 \left( 1 - 2 m(x) / x \right) } \: dx .
    \label{eq:3}
\end{equation}

%%%%%%%%%%%%%%%%%%%%%%%%%
\section{Matter content}
%%%%%%%%%%%%%%%%%%%%%%%%%

\subsection{Equation-of-state}

Thanks to the remarkable progress in observational astrophysics over the last few years, recent multimessenger observations have dramatically improved constraints on the equation-of-state of compact stars. The possibility that deconfined quark matter exists in the cores of compact stars—or that some compact stars are self-bound strange quark stars—has become increasingly relevant. Studying quark stars therefore provides a valuable framework for testing whether present observations favor purely hadronic matter, hybrid configurations, or absolutely stable strange quark matter. In this study we propose to model one of the compact stars shown in Table I of \cite{Aziz}. To be more precise, we model Cen X-3, the mass and radius of which take values in the range $1.41 \leq M/M_{\odot} \leq 1.57$ and $9.048 \leq R/km \leq 9.308$ \cite{citemass, citeradius}. 

\smallskip

Strange quark stars are based on the seminal works of Itoh \cite{Itoh:1970uw}, Bodmer \cite{Bodmer:1971we}, Terazawa \cite{Terazawa:1989iw} and Witten \cite{Witten:1984rs}, where it was proposed that strange quark matter consisting of up, down and strange quarks in weak equilibrium could replace $^{56}$Fe as the ground state of Quantum Chromodynamics at asymptotically large densities. According to this idea, the quarks in the stellar interior become effectively massless as compared with the associated chemical potential at very large densities, forming Cooper pairs with a common Fermi momentum. Since those pairs are electrically neutral, electrons cannot be present in this superfluid ground state \cite{Rajagopal:2000ff}, dubbed color-flavor locked (CFL) phase. The associated energy density and pressure at quadratic order in the s-quark mass, $m_s$, take the form in parametric form \cite{Lugones:2002va} 
\begin{eqnarray}
\rho(\mu) & = & \frac{9 \mu^4}{4 \pi^2 } - \frac{3\, m_s^2\, \mu^2}{4 \pi^2 } + \frac{3}{\pi^2} \Delta^2 \mu^2 + B_0\,,\hspace{10mm} \\
p(\mu) & = & \frac{3 \mu^4}{4 \pi^2 } - \frac{3 m_s^2 \mu^2}{4 \pi^2 }
+ \frac{3}{\pi^2} \Delta^2 \mu^2 - B_0
\end{eqnarray}
where the chemical potential, $\mu$, and the constant parameter $\alpha$ are given by
\begin{equation}
\mu^2 = - \alpha + \bigg( \alpha^2 + \frac{4}{9} \pi^2 (\rho - B_0) \bigg)^{1/2}\,, \hspace{10mm}
\alpha = -\frac{m^{2}_{s}}{6}+\frac{2\Delta^{2}}{3}\,,
\end{equation}
Finally, $\Delta$ is the superconducting gap and $B$ is a phenomenological bag constant encoding the difference between the ''perturbative vacuum" and the true vacuum.\footnote{In the MIT bag model \cite{Chodos:1974je,Chodos:1974pn,Farhi:1984qu}, hadrons consist of free or weakly interacting quarks confined to a finite region of space. This region or “bag” is stabilized by adding by hand a term $g_{\mu\nu} B$ to the energy-momentum tensor inside the bag.} Combining these equations, we get a pressure-density relation \cite{Lugones:2002va}
\begin{equation}\label{EoSstrange}
p= \frac13(\rho- 4B_0)
+ \frac{2 \Delta^2 \mu^2}{\pi^2} -  \frac{ m_s^2 \mu^2}{2 \pi^2 }\,.
\end{equation}
The first piece in this expression corresponds to the usual MIT bag model (radiation plus constant) for strange quark matter systems \cite{Chodos:1974je,Chodos:1974pn,Farhi:1984qu}. The second one, proportional to $\Delta^2$, is associated with the binding energy of the di-quark condensate and tends to make the system stiffer. The last term, proportional to $m_s^2$, has the opposite effect. 

\smallskip

Since the numerical values of the quantities $m_s$, $B$ and $\Delta$ characterizing the EoS \eqref{EoSstrange} are not accurately known, they will be considered as free parameters in this work. Among the many viable cases  within the stability conditions \cite{Flores:2017hpb}
\begin{equation}
m_s^2  <  2 \mu \Delta \,, \hspace{10mm}
B_0  < \frac{m_n^4}{108 \pi^2} + \frac{m_n^2 \Delta^2}{2 \pi^2} - \frac{m_s^2 m_n^2}{12 \pi^2}\,,
\end{equation}
with $m_n \simeq 939~{\rm MeV}$ the neutron mass, we shall consider here a scenario with $B_0=120\, \textrm{MeV}\cdot \textrm{fm}^{-3},$ $m_s=150\, \textrm{MeV}$ and $\Delta=150\, \textrm{MeV}$, in agreement with other phenomenological studies suggesting that $B_0 > 57~{\rm MeV}/{\rm fm}^3$ \cite{Farhi:1984qu} and $\Delta=(100-200)~{\rm MeV}$ \cite{Baym:2017whm}. This corresponds to the CFL18 model in Table I of Ref.~\cite{Flores:2017hpb}.

\smallskip

At this point let us justify the stability conditions of viable models and chosen parameters. In order for CFL quark matter to be absolutely stable, the energy per baryon must be lower than the neutron mass at zero pressure and temperature. Therefore, we must have $3 \mu < m_n$. Since this must hold at the zero pressure point, using the expression for the chemical potential we obtain the second condition. What is more, it has been shown in \cite{Kouvaris2004} that there is a quantum phase transition from the CFL phase to a new “gapless CFL phase”, and that the transition occurs where $m_s^2=2 \mu \Delta$.

\subsection{Anisotropic factor}

Recently the concept of complexity for self-gravitating systems within GR was introduced in \cite{herrera}. The so called complexity factor, which is a measure of complexity, appears in the orthogonal splitting of the Riemann tensor. Obviously, it vanishes for homogeneous energy densities and isotropic fluid spheres, but it may also vanish when the two terms containing density inhomogeneity and anisotropic pressure cancel each other, see for instance \cite{Greg1, comp1, comp2, comp3, comp4, comp5, comp6, comp7, comp8, comp9, comp10} for works on anisotropic stars within the complexity factor formalism.

In this context of the vanishing complexity factor, anisotropy is negative since the energy density is a decreasing function. 

In the present article we propose to incorporate anisotropies in three different ways as follows:
\\
\\
a) The Horvat ansatz, a purely phenomenological anisotropy model, according to which anisotropy is proportional to the radial pressure and the factor of compactness of the star \cite{Horvat} 
\begin{equation}
\Pi(r) = \kappa \left( \frac{2 m(r)}{r} \right) \: p_r(r)
\end{equation}
with $\kappa$ being a dimensionless coupling measuring the strength of the anisotropy. The case of relativistic stars made of isotropic matter is included in the limit $\kappa \rightarrow 0$. This form of anisotropy ensures that it vanishes at the center, and that anisotropies are negligible in the case of Newtonian stars. Motivated by Herrera's vanishing complexity factor, we shall consider in the following a negative coupling $\kappa=-0.5$.
\\
\\
b) Generate an exact analytic solution, as was done in \cite{Sharma} (albeit for a different EoS), assuming a certain radial profile for the mass function
\begin{equation}
m(r) = \frac{b r^3}{2 (1+a r^2)}
\end{equation}
which vanishes at the center of star as $m(r) \sim r^3$. The first TOV equation allows us to compute the energy density
\begin{equation}
\rho(r) = \frac{m'(r)}{4 \pi r^2} = \frac{b (3 + a r^2)}{(1+a r^2)^2}
\end{equation}
Next, the radial pressure is immediately computed via the adopted EoS. Finally, the anisotropic factor may be computed using the fluid equation
\begin{equation}
\Pi(r) = \frac{r}{2} \left[ p_r'(r) + (p_r(r) + \rho(r)) \frac{m(r) + 4 \pi r^3 p_r(r)}{ r^2 \left( 1 - 2 m(r) / r \right) } \right]
\end{equation}
The numerical values of the constant parameters $a,b$ may be computed using the matching conditions $p_r(R)=0, m(R)=M$. Considering $M=1.5~M_{\odot}$ and $R=9.1~km$, $a$ and $b$ are found to be 
\begin{equation}
a = \frac{8.67}{(30~km)^2}, \; \; \; \; \; b = \frac{9.51}{(30~km)^2}.
\end{equation}
\\
\\
c) For the sake of comparison, we shall also consider the case of Herrera's vanishing complexity factor, according to which the anisotropy is not arbitrarily imposed, but constrained by the internal gravitational structure
\begin{equation}
\Pi(r) = \frac{2}{r^3} \int_0^r dx \frac{1}{x^3} \rho'(x).
\end{equation}
That is conceptually attractive, since it reduces arbitrariness in model building, because it ties the matter sector directly to spacetime geometry. In this case instead of a certain EoS we shall assume the radial profile $m(r)$ of case b), and so we obtain an analytic expression for the factor of anisotropy
\begin{equation}
\Pi(r) = -\frac{a b \: r^2}{8 \pi (1+a r^2)^2},
\end{equation}
which is computed to be negative, and it vanishes at the center of the star.

%%%%%%%%%%%%%%%%%%%%%%%%%%%%%%%%%%%%%%%%%%%%%%%%%
\section{Radial oscillations of pulsating stars}
%%%%%%%%%%%%%%%%%%%%%%%%%%%%%%%%%%%%%%%%%%%%%%%%%

Considering a spherically symmetric system with only radial motion, Einstein's field equations can be used to compute the radial oscillation properties for a static equilibrium structure \cite{Chandrasekhar:1964zz, Chandrasekhar:1964zza, Kokkotas:2000up}. The radial perturbations of the stars are defined by \cite{chanmugam1, chanmugam2}
\begin{equation}
    \displaystyle \xi \equiv \frac{\Delta r}{r},
\end{equation}
\begin{equation}
    \displaystyle \eta \equiv \frac{\Delta p}{p}, 
\end{equation}

\medskip

\noindent where $\Delta r$ is the radial displacement and $\Delta p$ the pressure perturbation, which satisfy the following first order differential equations \cite{chanmugam1, chanmugam2}
\begin{align}
    \displaystyle \xi'(r) &= - \left( \frac{3}{r} + \frac{p'}{\zeta} \right) \xi - \frac{1}{r \Gamma} \eta , \\
    \displaystyle \eta'(r) &= \omega^2 \left[ r \left( 1 + \frac{\rho}{p} \right) e^{\lambda - \nu} \right] \xi - \left[ \frac{4 p'}{p} + 8 \pi \zeta r e^{\lambda} - \frac{r (p')^2}{p \zeta} \right] \xi - \left[ \frac{\rho p'}{p \zeta} + 4 \pi \zeta r e^{\lambda} \right] \eta , 
\end{align}

\medskip

\noindent where $e^{\lambda}$ and $e^{\nu}$ are the metric potentials shown in the equations \eqref{eq:2} and \eqref{eq:3}, respectively. Also, $\zeta$ is defined to be
\begin{equation}
    \displaystyle \zeta \equiv p + \rho . 
\end{equation}

\medskip

In addition, $\Gamma$ is the relativistic adiabatic index.
\begin{equation}
    \Gamma = c_s^2 \left( 1 + \frac{\rho}{p} \right) ,
\end{equation}

\medskip

\noindent with $c_s^{2}$ being the speed of sound squared given by
\begin{equation}
    c_s^2 = \frac{d p}{d \rho} . 
\end{equation}

\medskip

The equations for the perturbations contain singularities at the center and the surface. In order for the solutions to be regular everywhere, the coefficient of $1/r$ in the equation for $\xi$ must vanish as $r \rightarrow 0$. Similarly, the coefficient of $\rho/p$ in the equation for $\eta$ must vanish as $r \rightarrow R$ \cite{chanmugam1}. Hence, the unknown frequencies are determined solving the boundary value problem imposing the following conditions at the center and the surface of the star \cite{chanmugam1}
\begin{align}
    \displaystyle \frac{\eta}{\xi} \biggm|_{r = 0} &= - 3 \Gamma (0) , \\
    \displaystyle \frac{\eta}{\xi} \biggm|_{r = R} &= \left( 1 - \frac{2 M}{R} \right)^{-1} \left( - \frac{M}{R} - \frac{\omega^2 R^3}{M} \right) - 4  .
\end{align}

\medskip

The above equations for perturbations allow us to study the radial oscillation modes of stars made of isotropic matter. In the case of anisotropic stars, there are some additional terms that are proportional to the anisotropic factor, and therefore the linear system of coupled perturbations reads \cite{Arbanil}
\begin{equation}\label{ksi}
     \xi'(r) = -\frac{1}{r} \Biggl( 3\xi + \frac{\eta}{\Gamma} \Biggr) - \left( \frac{P'(r)}{P+\rho} + \frac{2 \Pi}{r P \Gamma} \right) \xi(r),
\end{equation}
\begin{equation}\label{eta}
    \begin{split}
          \eta'(r) = \xi \Biggl[ \omega^{2} r (1+\rho/P) e^{\lambda - \nu } - \frac{4P'(r)}{P} -8\pi (P+\rho) re^{\lambda} \frac{P+\Pi}{P} \\
     +  \frac{8 \Pi}{r P} + \frac{r(P'(r))^{2}}{P(P+\rho)}\Biggr] + \eta \Biggl[ -\frac{\rho P'(r)}{P(P+\rho)} -4\pi (P+\rho) re^{\lambda}\Biggr] ,
    \end{split}
\end{equation}

\medskip

\noindent where now $P,\Gamma$ are the radial quantities, i.e. $P(r)=p_r(r), \Gamma(r)=\Gamma_r(r)$.

\smallskip

The reason for appearing of the additional $\Pi$ terms is the following. According to the fluid equation (third TOV equation) a non-vanishing anisotropy behaves like an additional force. Moreover, when a shell of a pulsating anisotropic star oscillates $p_r$ and $p_t$ respond differently.

\smallskip

Finally, the frequency oscillation mode is computed by
\begin{equation}
\omega = s \: \omega_* , 
\end{equation}

\medskip 

\noindent where $s$ is a dimensionless number, while the constant $\omega_*$ is defined by
\begin{equation}
    \displaystyle \omega_* = \sqrt{\frac{M}{R^3}} .
\end{equation}

\medskip

Next, the frequencies are computed by
\begin{equation}
\nu_n = \frac{\omega_n}{2\pi}=\frac{s_n}{2 \pi} \:\sqrt{\frac{M}{R^3}} = s_n \frac{\sqrt{M/R}}{2\pi} \: \frac{1}{R},
\end{equation}

\medskip

\noindent where $n=0,1,2,...$ is the number of nodes for a star of a given mass and radius. Next, the conversion from geometrical units to $kHz$ is made using the conversion factors \cite{CamilaUniverse}
\begin{equation}
1 \: m = 5.068 \times 10^{15}~GeV^{-1}, \; \; \; \; \; 1 \: s = 1.519 \times 10^{24}~GeV^{-1},
\end{equation}
and finally $\nu_n=2.6 \: s_n~kHz$ in the case of the Cen X-3 for which $M/R=0.24$ and $R=9.1~km$.

\smallskip

A seismic parameter widely used in Asteroseismology is the so called large frequency separation, which is defined by
\begin{equation}
\Delta \nu_n = \nu_{n+1} - \nu_n, \; \; \; n=0,1,2,3, ...
\end{equation}

\medskip

\noindent and which can be easily computed once the spectrum is known. In other words, it is the difference between consecutive modes, and it can be shown to be related to stellar mass and radius \cite{tassoul, miglio, ilidio}.

\medskip

The condition $\omega_n^2=0$ separates the stable modes from the unstable ones. The frequency of the fundamental mode, $\nu_0$, at higher stellar masses is a decreasing function, and at some point vanishes. This marks the threshold of radial instability according to the Harrison-Zel’dovich criterion \cite{Harrison,Zeldovich}, see the discussion in Section 5.

\medskip

Before we present and discuss our main numerical results, let us comment in passing that although numerical methods and Ordinary Differential Equation solvers, such as Runge-Kutta, favor first order systems, it is advantageous to write down equivalently a master second order differential equation. Consider the system of two coupled first order differential equations of the form
\begin{eqnarray}
x'(r) & = & a x + b y \\
y'(r) & = & c x + d y    
\end{eqnarray}
where $a(r), b(r), c(r), d(r)$ are known functions of $r$. It is straightforward to manipulate both equations in order to write down the following second order differential equation
\begin{equation}
x''(r) + A(r) x'(r) + B(r) x(r) = 0,
\end{equation}
where the coefficients $A,B$ are computed in terms of $a,b,c,d$ as follows
\begin{eqnarray}
A & = & - \left( a + d + \frac{b'}{b} \right), \\
B & = &  a^2 + a  \frac{b'}{b} -a' + ad-bc.
\end{eqnarray}
There are several reasons for which it is worth studying a second order differential equation rather than a first order system. To mention just a few, i) access to Sturm-Liouville theory, ii) connection to well-known differential equations (such as Bessel, Airy, Hypergeometric, Heun etc), iii) an effective potential interpretation. Moreover, methods such as Frobenius, WKB, and Chebyshev spectral methods, are naturally formulated for second order equations.

%%%%%%%%%%%%%%%%%%%%%%%%%%%
\section{Numerical results}
%%%%%%%%%%%%%%%%%%%%%%%%%%%

Since the first order system of two coupled differential equations is linear, only the ratio $\eta/\xi$ matters. Therefore, without loss of generality we can impose the following conditions at the center of the star
\begin{equation}
\xi(0) = 1, \; \; \; \; \; \; \eta(0) = -3 \Gamma_r(0).
\end{equation}
For a given stellar mass and radius, after integrating the equations for the perturbations throughout the star, the values of the eigenfunctions at the surface of the object are a function of the unknown frequency only. To determine the allowed values of $\omega$ we have to solve graphically the algebraic equation (37). We use Wolfram Mathematica and its standard built-in commands, such as NDSolve, Plot, FindRoot etc. We did not face any convergence issues during the numerical computation.

\smallskip

\begin{figure}
\begin{center}
\includegraphics[scale=0.95]{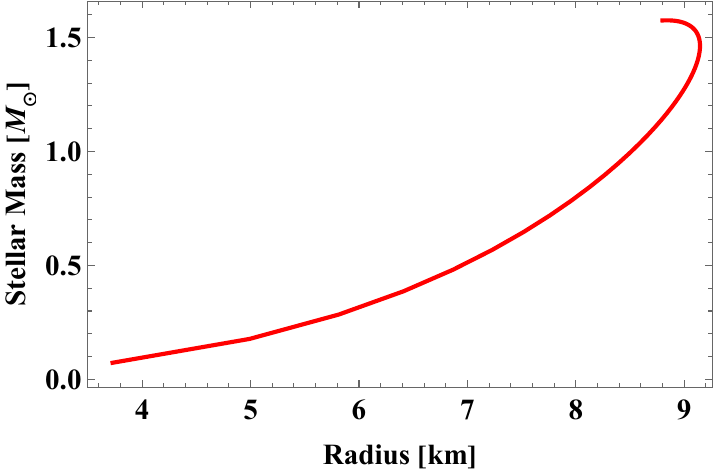} \
\includegraphics[scale=1]{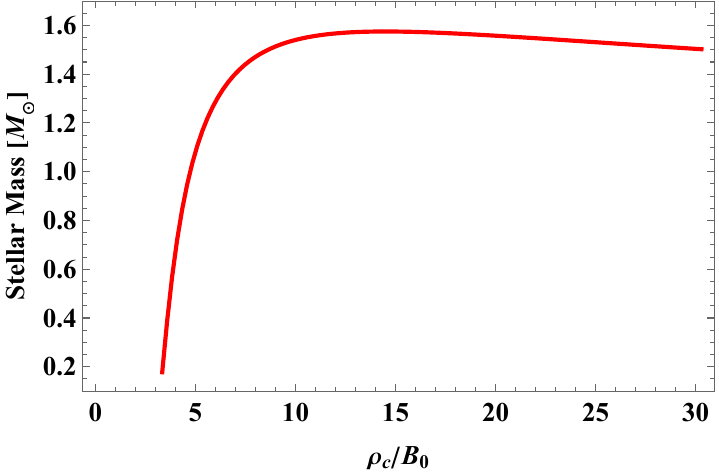} 
\caption{
{\bf Top panel:} Mass-to-radius relationship (radius in km and stellar mass in solar masses) in the case of model 1, see text. The highest stellar mass is computed to be $M_{max}=1.58~M_{\odot}$. {\bf Lower panel:} Stellar mass versus normalized central energy density (in units of the bag constant) in the case of model 1. The highest stellar mass is found to be $M_{max}=1.58~M_{\odot}$.
}
\label{fig0}
\end{center}
\end{figure}

We now present and discuss our main results displayed in the figures below, while the numerical values of the frequencies of the ten lowest modes for all three anisotropic models are shown in Table I. The mass-to-radius relationship in the case of the anisotropy model 1 is displayed in the top panel of Fig. \ref{fig0}. Our results show that the highest stellar masses that can be supported by the model considered here, namely CFL EoS plus the Horvat ansatz, is $M_{max}=1.58~M_{\odot}$. Furthermore, the lower panel of Fig. \ref{fig0} shows the relation between stellar mass and central energy density. According to the Harrison-Zel’dovich criterion \cite{Harrison,Zeldovich},
\begin{equation}
\frac{dM}{d \rho_c} > 0, \; \; \; \rightarrow  \; \; \; \textrm{stability},
\end{equation}
\begin{equation}
\frac{dM}{d \rho_c} < 0, \; \; \; \rightarrow  \; \; \; \textrm{instability},
\end{equation}
only the increasing part of the curve corresponds to stable configurations. The highest stellar mass observed here is the same maximum mass shown in the $M-R$ profile, and the same mass that marks the threshold of radial instability, $\nu_0^2=0$.

\smallskip

\begin{figure}
\begin{center}
\includegraphics[scale=0.9]{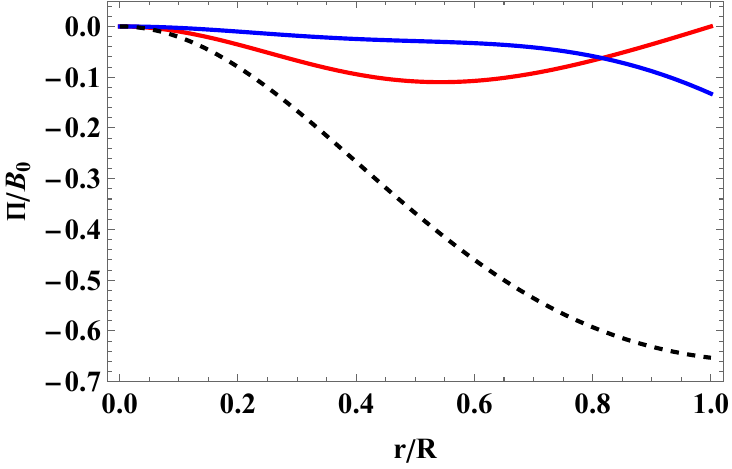} \
\includegraphics[scale=0.85]{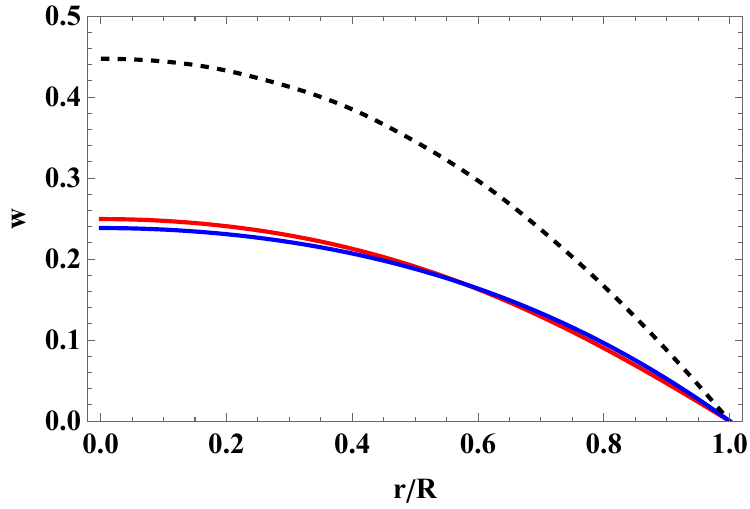}
\caption{
{\bf Upper panel:} Dimensionless anisotropic factor, $\Pi/B_0$, versus dimensionless radial coordinate, $x=r/R$, for the three anisotropy models discussed here, see text. The red curve corresponds to model 1, the blue curve to model 2, while the black dashed curve corresponds to Herrera's vanishing complexity factor. Only the anisotropic factor of model 1 vanishes both at the center and at the surface of the star. {\bf Lower panel:} Radial pressure-to-energy density ratio versus dimensionless $r$ for the three models discussed in this work. Since according to the matching conditions $p_r(R)=0$, the parameter $w$, too, vanishes at the surface of the star.
}
\label{fig1}
\end{center}
\end{figure}

In the two panels of Fig. \ref{fig1} we show the dimensionless anisotropic factor, $\Pi/B_0$ (in units of the bag constant), as well as the ratio $w=p_r/\rho$ as a function of the dimensionless radial coordinate, $x=r/R$ (in units of the stellar radius). The red solid curves are for model 1, blue solid curves for model 2 and the black dashed curves are for model 3. In all three cases the anisotropic factor is negative, although only the anisotropic factor of model 1 vanishes both at the center and at the surface of the star. The fact that $w$ is positive and lower than unity implies that the radial pressure is positive and lower than the energy density. Since according to the matching conditions $p_r(R)=0$, the parameter $w$, too, vanishes at the surface of the star. Furthermore, the curves corresponding to models 1 and 2 lie one close to another, whereas the curve corresponding to Herrera's method lies far apart from the other two.

\smallskip

\begin{figure}
\begin{center}
\includegraphics[scale=0.9]{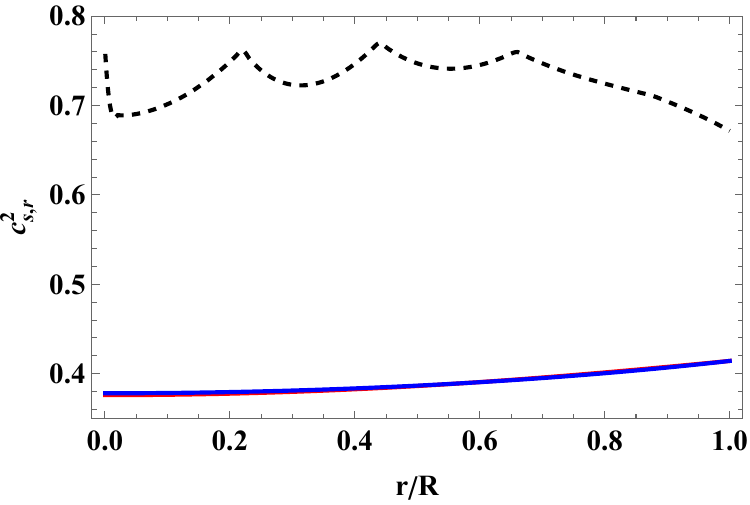} \
\includegraphics[scale=0.85]{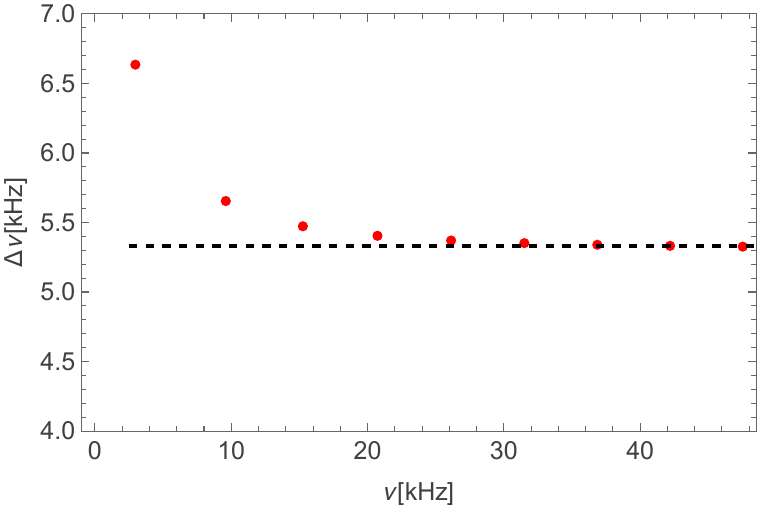}
\caption{
{\bf Upper panel:} Radial speed of sound versus radial coordinate for the three models discussed here. The red curve corresponds to model 1, the blue curve to model 2, while the black dashed curve corresponds to Herrera's vanishing complexity factor. {\bf Lower panel:} Large frequency separations versus frequency (both in kHz) for model 1. At higher excited modes the asymptotic value is found to be $5.33~kHz$.
}
\label{fig2}
\end{center}
\end{figure}

In the upper panel of Fig. \ref{fig2} we show the radial sound speed versus the dimensionless radial coordinate for all three anisotropic models discussed here. Colors are as in Fig. \ref{fig1}, and $c_{s,r}^2$ takes values in the range from 0 to unity. As in the previous figure the curves of the first two models lie one very close to another, whereas the curve of Herrera's approach lies further apart. Next, in the lower panel of \ref{fig2} we show the large frequency separations versus frequencies for model 1. The horizontal dashed line indicates the asymptotic value at higher excited modes. 

\smallskip

\begin{figure}
\begin{center}
\includegraphics[scale=0.85]{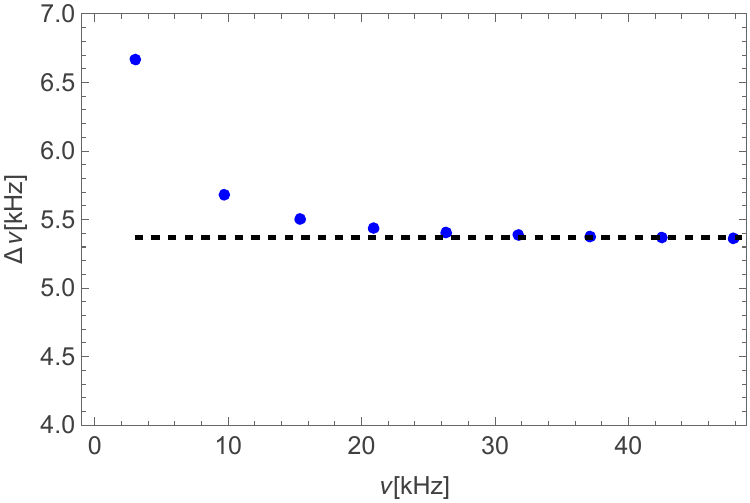} \
\includegraphics[scale=0.85]{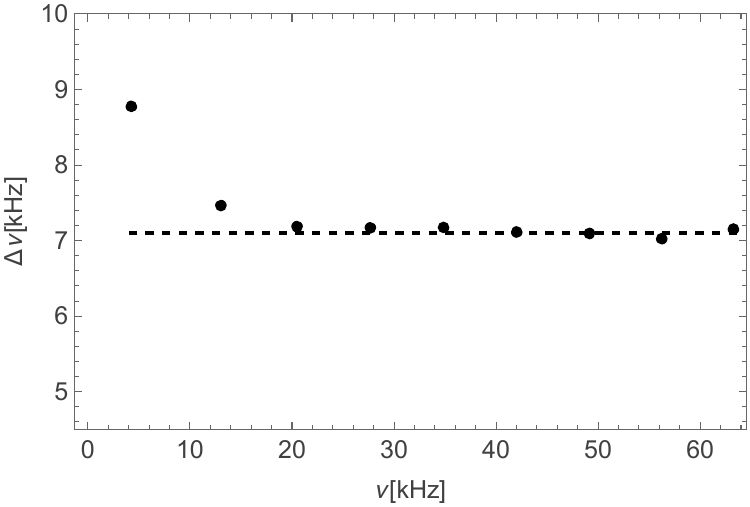}
\caption{
{\bf Upper panel:} Large frequency separations versus frequency (both in kHz) for model 2. At higher excited modes the asymptotic value is found to be $5.36~kHz$. {\bf Lower panel:} Large frequency separations versus frequency for model 3. At higher excited modes the asymptotic value is found to be $\sim 7.1~kHz$.
}
\label{fig3}
\end{center}
\end{figure}

Next, in the two panels of Fig. \ref{fig3} we show the large frequency separations versus frequencies for models 2 and 3. Same as before, the horizontal dashed lines indicate the asymptotic values at higher excited modes. It is observed that both the frequencies themselves and the large frequency separations increase from model 1 to model 3. The small oscillating feature around the asymptotic value observed in model 3 is due to the oscillating behaviour of the corresponding speed of sound. What is more, the difference between model 1 and 2 is small, whereas the difference between model 2 and 3 is significant. This is due to the fact that both the anisotropic factor and the sound speed of model 3 is very different compared to the other two models. Recall that model 3 was built up in a different manner compared to models 1 and 2, and so differences between oscillation spectra may arise due to anisotropies as well as model differences. To be more precise, comparing the first two models we observe a relative difference of $2.7 \%$ and $1.2 \%$ in the case of the first two modes, and a relative difference of $0.7 \%$ in the case of the last two modes. Moreover, comparing models 2 and 3 we observe a relative difference of $39.7 \%$ and $34.1 \%$ in the case of the first two modes, and a relative difference of $32.1 \%$ and $32.3 \%$ in the case of the last two modes.

\smallskip

\begin{figure}
\begin{center}
\includegraphics[scale=0.85]{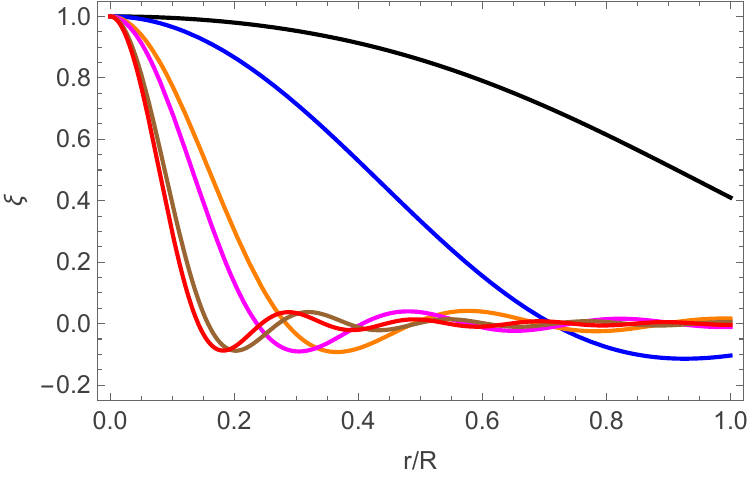} \
\includegraphics[scale=0.85]{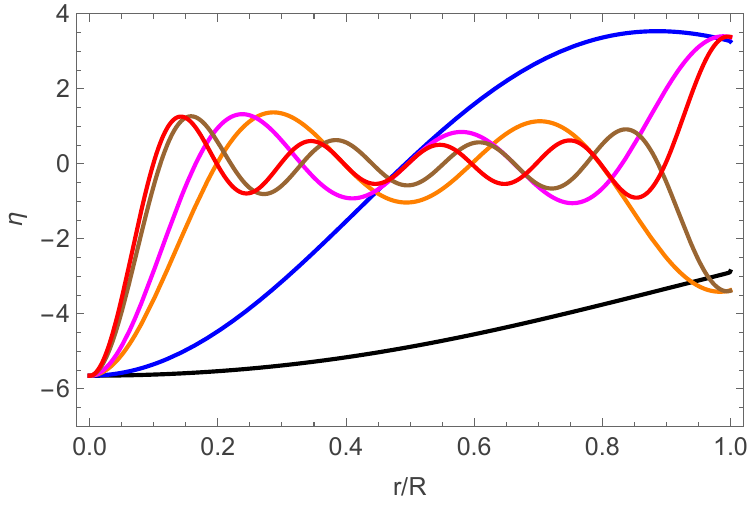}
\caption{
Eigenfunctions $\xi, \eta$ versus radial coordinate for model 1. Shown are $n=0$ (black), $n=1$ (blue), $n=4$ (orange), $n=5$ (magenta), $n=8$ (brown) and $n=9$ (red).
}
\label{fig4}
\end{center}
\end{figure}

Finally, the radial profiles of the eigenfunctions $\xi,\eta$ are shown in Fig. \ref{fig4}. Here we have included the case of model 1 only, since the pattern is almost identical in all cases. It is observed the typical behaviour seen in other studies, see for instance \cite{Greg2,Ishfaq,Camila}.
We have shown $\xi(r), \eta(r)$ corresponding to the first two excited modes ($n=0$ and $n=1$), two intermediate modes ($n=4$ and $n=5$) as well as the highest modes computed here ($n=8$ and $n=9$). At the origin the function $\xi(r)$ starts at unity, whereas the function $\eta(r)$ starts at $-3 \Gamma_r(0)$, as was imposed by the boundary condition (48). As expected, the number of nodes is equal to the overtone number $n$. In other words, as in every Sturm-Liouville boundary value problem, the eigenfunctions of the fundamental mode have no zeros, the ones of the first excited mode have a single node, and so on and so forth.

\smallskip

Before we conclude and summarize our work, a couple of comments are in order regarding limitations and future work. First, conclusions were drawn from a single model only as all computations were made for $M=1.5 \: M_{\odot}$ and $R=9.1 \: km$. A sensitivity analysis on the coupling $\kappa$, and a simple parameter scan within the observational mass-radius range would strengthen the conclusions. Additionally, while the present work was restricted to radial oscillations, which do not emit gravitational radiation in spherical symmetry, the significant changes in the radial-mode spectrum found for different anisotropy models motivate future studies of non-radial oscillations. Such modes are directly relevant for gravitational-wave asteroseismology, and may provide an additional observational channel for testing anisotropic stellar models. We hope to be able to investigate in detail those important issues in the near future.

%%%%%%%%%%%%%%%%%%%%%%%%TABLE%%%%%%%%%%%%%%%%%%%%%%%%%%%

\begin{table}[]
\centering
\caption{
Frequencies (in kHz) for the three anisotropic models considered here. 
}
\label{tab:frequencies}
{
\resizebox{0.4\columnwidth}{!}
{
\begin{tabular}{@{}c|ccc@{}}
\toprule
\multicolumn{4}{c}{Frequencies at different mode orders}            \\ 
\hline
\hline
%\midrule
$n$ & $\textrm{Model 1}$  &  $\textrm{Model 2}$  &  $\textrm{Model 3}$ \\
\hline
\hline
0   & 2.966   & 3.046   &  4.254  \\
\hline
\hline
1   & 9.601   & 9.714   & 13.031  \\
\hline
\hline
2   & 15.256  & 15.395  & 20.494  \\
\hline
\hline
3   & 20.728  & 20.900  & 27.679  \\
\hline
\hline
4   & 26.133  & 26.338  & 34.848  \\
\hline
\hline
5   & 31.503  & 31.744  & 42.023  \\
\hline
\hline
6   & 36.855  & 37.131  & 49.137  \\
\hline
\hline
7   & 42.195  & 42.508  & 56.232  \\
\hline
\hline
8   & 47.527  & 47.877  & 63.259  \\
\hline
\hline
9   & 52.853  & 53.240  & 70.413
\end{tabular}
}
}
\end{table}

%%%%%%%%%%%%%%%%%%%%%%%%%%%%%%%%%%%%%%%

%%%%%%%%%%%%%%%%%%%%%%%%%%%%PLOTS%%%%%%%%%%%%%%%%%%%%%%%%%%%

%%%%%%%%%%%%%%%%%

%%%%%%%%%%%%%%%%%

%%%%%%%%%%%%%%%%%%%%%%%

%%%%%%%%%%%%%%%%%%%%%%%%%%%

%%%%%%%%%%%%%%%%%%%%%END-OF-PLOTS%%%%%%%%%%%%%%%%%%

%%%%%%%%%%%%%%%%%%%%%%%%%%%%%%%%%
\section{Summary and conclusions}
%%%%%%%%%%%%%%%%%%%%%%%%%%%%%%%%%

To summarize our work, in the present article we computed the spectra of radial oscillations modes of pulsating relativistic stars made of anisotropic matter within four-dimensional Einstein's gravity without a cosmological constant. The star was modeled as a fluid sphere characterized by stellar mass and radius without rotation and without a net electric charge. The anisotropic factor was introduced in three different ways, namely the purely phenomenological Horvat model, where the factor of anisotropy is proportional to the radial pressure and the factor of compactness (model 1), an exact analytic solution assuming a certain radial profile for the mass function (model 2), and Herrera's vanishing complexity factor (model 3) modeling the same compact object of mass $M=1.5~M_{\odot}$ and radius $R=9.1~km$. Contrary to the models 1 and 2, where the CFL analytic EoS for quark matter was assumed, model 3 does not require a certain EoS.

\smallskip

First the structure equations describing hydrostatic equilibrium of interior solutions were presented. Next, the details of matter content regarding the underlying EoS and factor of anisotropy were specified. The equations for the perturbations of pulsating stars were also briefly discussed. The numerical values of the frequencies of the ten lowest radial oscillation modes were reported in Table I, while the normalized anisotropy, the radial sound speed and the pressure-to-energy density ratio for all three models were displayed in three panels of Figures \ref{fig1} and \ref{fig2}. Moreover, the large frequency separations were shown in three panels of Figures \ref{fig2} and \ref{fig3}. In all three cases considered here, the factor of anisotropy is negative, while the large frequency separations tend to an asymptotic value at highly excited modes. What is more, the frequencies of the modes corresponding to model 1 were found to be the lowest, whereas the frequencies of the modes corresponding to Herrera's approach were computed to be the highest. Finally, the sound speeds and the pressure-to-density ratios of models 1 and 2 lie one very close to another, whereas the curves corresponding to Herrera's method are observed to be significantly different.

\smallskip

Although the present work is focused on the theoretical properties of radial oscillations in anisotropic compact stars, it is worth commenting on the possible observational implications of the obtained frequency spectrum. The oscillation frequencies depend sensitively on the internal stellar structure and on the adopted anisotropy prescription. In our calculations, the first two anisotropy models produce relatively small differences in the eigenfrequencies (typically of the order of (1-2)$\%$, whereas the third model predicts substantially larger deviations, reaching approximately (30-40)$\%$ for the stellar configurations considered. Although radial oscillations do not generate gravitational waves in spherical symmetry, they provide valuable information on stellar stability and the underlying microphysics. Future developments in compact-star asteroseismology, combined with increasingly accurate modeling of stellar oscillations, may therefore offer indirect constraints on the anisotropy of dense matter. A quantitative analysis of the observational detectability of the predicted frequency differences is beyond the scope of the present work.

\section{Acknowledgments}
The author wishes to thank the anonymous referees for useful comments and~suggestions.

% If authors have biography, please use the format below
%\section*{Short Biography of Authors}
%\bio
%{\raisebox{-0.35cm}{\includegraphics[width=3.5cm,height=5.3cm,clip,keepaspectratio]{Definitions/author1.pdf}}}
%{\textbf{Firstname Lastname} Biography of first author}
%
%\bio
%{\raisebox{-0.35cm}{\includegraphics[width=3.5cm,height=5.3cm,clip,keepaspectratio]{Definitions/author2.jpg}}}
%{\textbf{Firstname Lastname} Biography of second author}

% For the MDPI journals use author-date citation, please follow the formatting guidelines on http://www.mdpi.com/authors/references
% To cite two works by the same author: \citeauthor{ref-journal-1a} (\citeyear{ref-journal-1a}, \citeyear{ref-journal-1b}). This produces: Whittaker (1967, 1975)
% To cite two works by the same author with specific pages: \citeauthor{ref-journal-3a} (\citeyear{ref-journal-3a}, p. 328; \citeyear{ref-journal-3b}, p.475). This produces: Wong (1999, p. 328; 2000, p. 475)

%%%%%%%%%%%%%%%%%%%%%%%%%%%%%%%%%%%%%%%%%%
%% for journal Sci
%\reviewreports{\\
%Reviewer 1 comments and authors’ response\\
%Reviewer 2 comments and authors’ response\\
%Reviewer 3 comments and authors’ response
%}
%%%%%%%%%%%%%%%%%%%%%%%%%%%%%%%%%%%%%%%%%%
%\end{adjustwidth}

\end{document}